\documentclass[acmsmall]{acmart}

\usepackage{xcolor}
\usepackage{hyperref}
\usepackage{graphicx}
\usepackage{subcaption}
\usepackage{booktabs}
\usepackage{caption}

\AtBeginDocument{%
  }

\setcopyright{acmlicensed}
\acmJournal{PACMHCI}
\acmYear{2025} \acmVolume{9} \acmNumber{2} \acmArticle{CSCW195} \acmMonth{4}\acmDOI{10.1145/3711093}

\begin{document}

\title{Translating Emotions to Annotations: A Participant's Perspective of Physiological Emotion Data Collection}


\author{Pragya Singh}
\orcid{0000-0003-3933-2224}
\affiliation{%
  \institution{IIIT-Delhi}
  \city{New Delhi}
  \country{India}}
\email{pragyas@iiitd.ac.in}

\author{Ritvik Budhiraja}
\orcid{0009-0004-2392-5441}
\affiliation{%
\institution{IIIT-Delhi}
  \city{New Delhi}
  \country{India}}
\email{ritvik193222@iiitd.ac.in}

\author{Pankaj Jalote}
\orcid{0009-0001-8552-8394}
\affiliation{%
\institution{IIIT-Delhi}
  \city{New Delhi}
  \country{India}}
\email{jalote@iiitd.ac.in}

\author{Mohan Kumar}
\orcid{0000-0002-0286-6997}
\affiliation{%
\institution{RIT}
  \city{Rochester}
  \country{New York, US}}
\email{mjkvcs@rit.edu}

\author{Pushpendra Singh}
\orcid{0000-0003-2152-1027}
\affiliation{%
\institution{IIIT-Delhi}
  \city{New Delhi}
  \country{India}}
\email{psingh@iiitd.ac.in}

\renewcommand{\shortauthors}{Pragya Singh, Ritvik Budhiraja, Pankaj Jalote, Mohan Kumar, \& Pushpendra Singh}

\begin{abstract}
  Physiological signals hold immense potential for ubiquitous emotion monitoring, presenting numerous applications in emotion recognition. However, harnessing this potential is hindered by significant challenges, particularly in the collection of annotations that align with physiological changes since the process hinges heavily on human participants. In this work, we set out to study human participants' perspectives in the emotion data collection procedure. We conducted a lab-based emotion data collection study with 37 participants using 360{$^\circ$} virtual reality video stimulus followed by semi-structured interviews with the study participants. Our findings presented that intrinsic factors like participants' perception, experiment design nuances, and experiment setup suitability impact their emotional response and annotation within lab settings. Drawing from our findings and prior research, we propose recommendations for incorporating participants' context into annotations and emphasizing participant-centric experiment designs. Furthermore, we explore current emotion data collection practices followed by AI practitioners and offer insights for future contributions leveraging physiological emotion data.
\end{abstract}





\maketitle

\section{Introduction}
Effective development and operation of Artificial Intelligence (AI) technologies hinge on the quality of data used. The impact of subpar data on the performance of AI models highlights the necessity for a thorough investigation into this aspect. Researchers in Computer Supported Cooperative Work (CSCW) and Human-Computer Interaction (HCI) have emphasized the significance of data work. They advocate for incentivizing efforts to enhance data quality and have stressed the necessity for transparent collaboration among stakeholders involved in data work and its implications on data quality \cite{10.1145/3290605.3300830, 10.1145/3274345, 10.1145/3449205, 10.1145/3432955, 10.1145/3411764.3445518, DBLP:journals/corr/abs-2001-06684}. Data work has emerged as crucial in building AI applications for critical domains such as healthcare \cite{10.1145/3411764.3445518}, Environmental Monitoring \cite{li2021ai}, and Education \cite{pechenkina2023artificial}.
While acknowledging the significance of data quality is essential, obtaining high-quality and well-annotated datasets poses its own challenges. These challenges are further accentuated when the data collection involves human participants as crucial stakeholders, which is frequently encountered in healthcare data scenarios.

Simultaneously the rise in wearable devices equipped with sensors for capturing physiological data has prompted a surge in research initiatives at the intersection of wearables, physiological signal data and AI, aimed explicitly towards recognizing emotions and monitoring mental disorders and well-being \cite{gjoreski2016continuous, can2023approaches, mishra2021detecting, 10.1145/3596246, 10.1145/3544548.3581209}. 
Like other healthcare domains, emotion recognition using physiological signal data and wearable devices relies heavily on human participants as significant stakeholders for data collection and labeling. Thus, the quality of physiological signal-based emotion data is also dependent on human participants. Moreover, other essential data quality parameters include using high-resolution wearable sensors with minimal noise or artifacts for physiological signal data collection, employing reliable methods for labeling emotional states, collecting data from diverse groups of participants, using realistic elicitation methods to collect emotional data that accurately reflect genuine emotional responses, and ensuring the data is complete with information on participants and their characteristics (such as age, gender, personality and health history) \cite{larradet2020toward, 9779458, singh2024saycatcatunderstanding, can2023approaches, bota2019review}.

Prior studies on collecting physiological emotion data have primarily adopted two approaches. The first involves intentionally eliciting specific emotions of interest in participants, while the second revolves around annotating emotions in real-life situations based on individuals' experiences \cite{9779458}. These approaches have been explored across three main settings: i) \emph{Laboratory Settings} \cite{schmidt_introducing_2018, miranda-correa_amigos_2021}, ii) \emph{Field with Constraints Settings} \cite{koldijk_swell_2014, hosseini_multimodal_2022}, and iii) \emph{Field Settings} \cite{mundnich_tiles-2018_2020, yfantidou_lifesnaps_2022, kang_k-emophone_2023}. However, each setting has its own set of limitations \cite{9779458}. For instance, models trained on data from laboratory settings often exhibit sub-optimal performance when deployed in real-world scenarios. Data collected in a field with constrained settings may be overly tailored to specific tasks, limiting its broader applicability. Similarly, due to limited environmental control, field settings face challenges such as imbalanced data, noise interference, and missing annotations.

Apart from experimenting with various settings, prior research has extensively examined various techniques for emotion elicitation, stimulus types, experimental methodologies, physiological sensors, and annotation methods \cite{9779458, bota2019review, can2023approaches}. 
However, a noticeable gap exists in the literature on emotion recognition investigating the impact of human participants in physiological emotion data collection procedures. Previous works have often given more importance to design decisions like the type of stimuli, the content of stimuli, or the data collection settings. However, limited attention is being given to what human participants' characteristics may impact their emotion elicitation, and thus the quality of data \cite{miranda-correa_amigos_2021, subramanian_ascertain_2018, tabbaa_vreed_2022, schmidt_introducing_2018}. In the past, CSCW and HCI researchers have repeatedly highlighted the importance of communication among all stakeholders for computing technology.
Researchers have also extensively explored the perspectives of human participants in AI data collection practices in various critical domains and have highlighted the detrimental effects that conventional AI/ML data collection and preparation methods can have on overall data quality \cite{10.1145/3411764.3445518, 10.1145/3491102.3501868}. Further recommendations were provided for establishing robust procedures based on insights from stakeholders in fields like healthcare and computer vision \cite{10.1145/3491102.3501868, 10.1145/3274345, andrews2023ethical}. However, participants' perspectives still need to be explored in the emotion recognition community.

This study aims to bridge this research gap and delve into human participants' perspectives regarding the physiological emotion data collection procedure.
Our focus is on exploring the following research questions:
\\

\textbf{RQ1:} \emph{What participant-specific factors can impact emotion elicitation, annotation, and data quality?}

\textbf{RQ2:} \emph{What are the participants' perspectives on the physiological emotion data collection process, including the labeling method, stimuli, and the experimental setup (VR)?}\\

We conducted a lab-based physiological emotion data collection experiment to address our research questions, followed by semi-structured qualitative interviews with 37 participants (21 males, 15 females, 1 undisclosed). Our choice of a lab setting aligns with the prevalent use of such settings in prior research on physiological emotion data collection \cite{miranda-correa_amigos_2021, subramanian_ascertain_2018, tabbaa_vreed_2022, schmidt_introducing_2018}. We utilized Virtual Reality (VR) short 360{$^\circ$} videos as stimuli within the lab environment. The selection of VR-based stimuli stems from their demonstrated ability to evoke a higher emotional response compared to 2D stimuli \cite{tian2021emotional, gilpin2021physiological}.
The following sections delve into related literature, our study design, and our findings. Next, we present a discussion on the role of participants, focusing on how their perceptions and interpretations influence their emotional responses and annotations. We then provide recommendations for practitioners and researchers in the CSCW, HCI, and AI communities to work towards collaborative physiological emotion data collection practices while keeping all stakeholders in mind.

\section{Background and Related Work}

Emotions have been a topic of study for many decades. For understanding and identifying emotions, several theories have emerged over a period of time to define emotions\cite{tian2022applied}. The theories of emotions can be categorized into Evolutionary, Cognitive Appraisal, and Social Constructivist theories. The Evolutionary theories \cite{plutchik1982psychoevolutionary} consider emotions as a product of evolution that are universal and can be categorized into distinct categories such as happy, sad, angry, etc. The Cognitive Appraisal theory of emotions \cite{roseman2001appraisal} proposes that emotions result from the cognitive interpretation of situations. Lastly, the Social Constructivist theory of emotions \cite{barrett2017theory} suggests emotions to be a construction of the brain based on cultural influences, past experiences, and the context of the situation and are an interplay of cognition, biology, culture, and current environment context and thus are subjective as opposed to previous beliefs of emotions being universal. Researchers in emotion recognition have used one or more of these emotional theories to label physiological emotion data. The circumplex model of emotion \cite{russell_affect_1989} is one of the most commonly used appraisal theory-based models in the past for labeling emotions in data collection. It defines emotions in various dimensions, namely, valence (the degree of positivity and negativity of emotion), arousal (the intensity of emotions from low energy to high energy), and dominance (the degree of control over emotions). To further ease the process of labeling, pictorial scales based on the circumplex model are also explored, like Self-Assessment Manikin (SAM) \cite{bynion2020self}, Geneva Emotion Wheel (GEW) \cite{shuman2015geneva}, and EmojiGrid \cite{toet2018emojigrid}. 
Further emotions are also labeled using evolutionary approaches using self-reporting questionnaires with a list of emotions such as Positive and Negative Affect Schedule (PANAS) \cite{watson_development_1988}, Emotions twenty-questions \cite{kazemzadeh2011emotion}, and Basic emotion theory \cite{ekman1992there}. The theories of emotions have been an essential set of guidelines for researchers collecting physiological emotion data labels. However, the qualitative experiences of human participants in emotional data collection need further investigation. This exploration is essential to understanding all stakeholders in emotion data work and improving the quality of data collection practices.

\subsection{Data-Work and Emotion Data}

Recognizing AI's pervasive influence across various sectors, recent research efforts have increasingly emphasized the need for more work to ensure data quality. While traditional AI communities predominantly prioritized model development, the significance of data work and transparent communication and collaboration among all stakeholders involved in the data collection procedure has now come to the forefront. As emphasized by Nithya et al. \cite{10.1145/3411764.3445518}, neglecting data quality during the initial stages of machine learning can result in compounded challenges later on. The emergence of data-centric AI as a pivotal concept has garnered attention from the research community and AI practitioners \cite{zha2023datacentric}. 
For maintaining data quality, prior machine learning literature has proposed metrics such as data completeness, label reliability, data integrity (data should follow realistic constraints), data consistency, accuracy, and amount of errors or noise \cite{gudivada2017data, 10.1145/3592616}. Further stage-wise data quality requirements for ML pipeline have been proposed for practical usage \cite{10.1145/3592616}. Moreover, with the rise in research on fairness and bias, data is also being assessed for fairness and bias metrics \cite{pitoura2020social}. The data collection pipeline has been studied at the process level to avoid data-related issues \cite{DBLP:journals/corr/abs-2107-01824, 10.1145/3351095.3372862}. Recent works in CSCW and HCI communities increasingly investigated the human infrastructure involved in data collection and labeling \cite{10.1145/3491102.3501868, 10.1145/3290605.3300830}. Human-centric approaches around data and decision-making are also emerging, suggesting a need for attention to human participants' involvement within data work \cite{10.1145/3491101.3516403, 10.1145/3449084, 10.1145/3359206}. 

Data quality parameters within physiological emotion data can be grouped into two major subgroups- 1) \textbf{Methodological Parameters}, which includes realistic or non-intrusive data collection settings (lab or real), emotionally engaging emotion elicitation methods (stimuli-based or daily life task-based), experiment setup (sensors and acquisition device), and controlled data collection environment (temperature, noise), and 2) \textbf{Data Labelling Parameters}, which includes how the data is labeled using what questionnaires or scales, when is data labeled (after how much duration), what other contextual data is collected to help the labeled data (information on food intake, caffeine intake, medical history etc.) and how accurately are labels defining the emotions experienced by participants. 
Prior work has explored methodological aspects, such as selecting appropriate stimuli for accurate emotional engagement, the choice of data collection settings, and experiment setups \cite{stemmler2003methodological, larradet2020toward}. Various stimuli, including pictures, films, music, games, virtual videos, and psychological and cognitive tasks, have been investigated for their effectiveness in emotion elicitation with different stimuli providing different levels of emotional engagement. Virtual videos were shown to have a high level of presence and immersion, while games have been shown to have a high level of interactivity \cite{coan2007handbook, uhrig2016emotion, bouga2023comparison, ellis2005impact, rivu2021emotion}.
The order and duration of stimuli have also been examined, with random stimulus order being commonly used to avoid any impact of the stimulus order, and both long and short stimuli explored based on experiment duration and study fatigue \cite{subramanian_ascertain_2018, saganowski_emognition_2022, miranda-correa_amigos_2021}. Longer stimuli have advantages such as deeper emotional engagement and elicit complex emotions that take time to develop. Shorter stimuli were used for less time-consuming, distinct emotions, which led to less fatigue in participants during experiments. The impact of the arrangement of stimuli was also studied in the past \cite{goshvarpour2015affective}, suggesting a relationship between stimuli sequence and emotional response. Prior work has also explored active stimuli like VR games \cite{9792298} and interactive storytelling applications \cite{weis_inducing_2022}, integrating physiological feedback \cite{10.1145/3242671.3242676}, and haptics-based immersive experiences to collect physiological emotion data in lab settings, offering enhanced control over experiment design \cite{10.1145/2929490.2932629}.
Experiment setup choices, including the choice of sensors (portable versus laboratory) \cite{ragot2018emotion}, type of sensors (wrist-worn or chest-worn etc.) \cite{schmidt_introducing_2018}, and the placement of sensors on the human body \cite{10.1145/3643541, 10.1145/3411764.3445370}, are scrutinized for their impact on physiological data. Picard et al. has elaborated on \textit{"Gathering good affective data"} \cite{954607}, highlighting the significance of sensor placements, minimizing artifacts, using gel before sensor placement, and accurate gathering of ground truth data for collecting high-quality data. They also identified five key factors as essential decisions for physiological emotion data collection: (i) how the emotion is elicited (subject or event), (ii) the setting of the experiment (lab or real), (iii) the emphasis of the data (external expressions or internal feelings), (iv) how the data is recorded (openly or hidden from the subject's knowledge), and (v) whether the subject knows the purpose of the experiment (emotion elicitation or otherwise).
Additionally, prior work has highlighted that physiological signals—such as heart rate, skin conductance, and brain activity—are inherently influenced by several participant-specific characteristics. These characteristics can significantly affect the interpretation and reliability of the data collected during emotion recognition studies. Some of the key participant-specific factors include- 1) Age, which is shown to impact the baseline physiological measures \cite{lin2021differences}, 2) Gender, which can influence the physiological measures due to the change in gender-specific emotional responses and hormonal fluctuations \cite{lambrecht2014gender, 9187875}, 3) Physical Fitness and Health, which may exhibit different baseline and reactive physiological responses and health issues that need to be accounted for to avoid confounding effects in data interpretation \cite{lin2021differences}, 4) Emotional well-being, which may exhibit heightened or dampened physiological responses to emotional stimuli complicating the interpretation of the data \cite{mcrae2017biological}, 5) Personality Traits that may cause different emotional responses influencing the physiological data recorded during emotional tasks \cite{subramanian_ascertain_2018}, and 6) Past experiences resulting in stronger or lack of physiological reactions to same stimuli \cite{barrett2017emotions}. To overcome these variations in data, prior work has collected physical activity \cite{gjoreski2017monitoring, 10.1145/3463508}, personality traits \cite{zhao2018personality}, and other participant characteristics as part of their datasets. However, there remains a scarcity of work utilizing contextual information for emotion recognition.

Further, for data labeling parameters, prior works have explored various labeling techniques \cite{singh2024saycatcatunderstanding}. The component process model and appraisal theory models have been utilized for annotating emotions \cite{mohammadi2020multi, meuleman2018induction} along with evolutionary approaches for accurately annotating emotions. Dimensions such as the significance of the situation \cite{imbir2016affective} and approach-avoidance motivation \cite{saganowski_emognition_2022} were also explored to add contextual information such as motivation and significance to annotations.
Schmidt et al. \cite{10.1145/3267305.3267551} proposed practical guidelines to improve the accuracy of labels in emotion data collection. 
Yang et al., in their work \textit{"Annotations 
matters"} have highlighted the difference in recognition performance of three different labeling methods, intended (as per stimuli-task), self-assessed (as assessed by participants), observed (as assessed by the external observer) \cite{yang2019annotation}, and found that predictability from self-assessed labeling data is worst given the uncertainty added by participant's perceptions and interpretation of emotions. Emotion Annotation is often identified as a problem with several layers \cite{6197603, singh2024saycatcatunderstanding} like, lack of understanding of what is ground truth, lack of conclusion on which method of annotation is accurate, and lack of a method to accurately capture emotions without bias.
Emotion data is prone to several human biases such as cognitive, motivational, and perception biases \cite{pronin2002bias, pronin2007perception, haselton2009adaptive, kruglanski1983bias}, which can significantly impact the accuracy of data, particularly during the labeling process. Emotions are complex psychological states involving three distinct components majorly: a subjective experience, a behavioral or expressive response, and a physiological response. The traditional methods of self-annotating emotions in past studies \cite{schmidt_introducing_2018, subramanian_ascertain_2018, miranda-correa_amigos_2021, saganowski_emognition_2022} often relied on objective scales or questionnaires that present a limited set of emotions. This approach may fail to capture the full range of emotions felt by participants, given their complexity and the reliance on human interpretation within the constraints of an experimental setup \cite{folz2022reading}.
Previous research has primarily concentrated on various aspects such as stimulus selection, collection environment, elicitation methods, labeling methods, and the acquisition of extensive physiological signals and contextual data through sensors and additional questionnaires. In prior work, participants' perspectives are often collected in the form of their self-reports, participant characteristics details and using contextual data like acceleration and participant's location and activity by using additional questionnaires \cite{gjoreski2017monitoring, hosseini_multimodal_2022}. However, there is a notable gap in the literature studying participants' experiences and perceptions qualitatively better to understand their overall procedure of emotion interpretation and labeling.
Social scientists have long employed hermeneutic approaches to interpret and understand human experiences. Ricoeur's theory on the "hermeneutics of the self" \cite{ricoeur1975philosophical} emphasizes the interpretative process through which individuals comprehend their own experiences and narratives. This theory acknowledges that individuals are not merely passive recipients of experiences but are active interpreters who construct meaning from their emotions and life events. Such theories are instrumental in understanding the subjective responses within emotion data collection procedures. 
This study aims to build upon hermeneutic theory from social science and integrate insights from previous data work across CSCW, HCI, and AI communities. Since emotions are subjective constructions correlated to physiological and behavioral changes, they require a complex interpretation process from the participant's side that is often not captured in objective scales alone.   
By conducting this study, we seek to understand more about the perspectives of human participants' qualitative interpretations while engaging in emotion data collection procedures within lab settings. 

\subsection{Emotion Data Collection Practices}

Emotion recognition, a rapidly growing field within AI and HCI, focuses on identifying and interpreting human emotions through various data modalities such as facial expressions, voice tone, physiological signals, and behavioural patterns. With the availability of wearable sensors, there is an increase in ongoing work in using physiological signals for emotion recognition \cite{siirtola_continuous_2019, 10.1145/3361562}. Physiological signals, unlike other modalities like facial expression, text, voice, and posture, are direct indicators of internal physiological processes and are less prone to conscious control and intentional masking of emotions. Prior works in this field have explored various types of stimuli, annotation methods, and biosignals for emotion recognition.

\textbf{Stimuli}: Within lab settings, emotion elicitation methods are used for recording an individual's emotional response to a stimulus \cite{saganowski_emognition_2022, kutt_biraffe2_2022, schmidt_introducing_2018, miranda-correa_amigos_2021, subramanian_ascertain_2018, singheevr}. Previous works have explored various methods of emotion elicitation or induction, such as the Induction through Dyadic interactions \cite{roberts2007emotion}, induction through images \cite{lang2005international}, induction through sounds \cite{stevenson2008affective, bradley2007international}, induction through music \cite{fuentes2021emotion}, induction through films \cite{gross1995emotion, baveye_affective_2018}, induction through games \cite{yannakakis2014emotion}, induction through haptic expressions \cite{gaffary_haptic_2020}, induction through psychological stress stimuli \cite{schmidt_introducing_2018} and induction through VR experiences \cite{li2017public}. While field settings have utilized daily-life stimuli for collecting physiological data \cite{kang_k-emophone_2023}.

\textbf{Emotion Annotations}: For annotating the physiological data, prior works have employed techniques such as annotations by external annotators or experts \cite{healey_detecting_2005}, self-reports using self-assessment questionnaire such as SAM or PANAS \cite{schmidt_introducing_2018}, experience sampling \cite{shui2021dataset}, continuous annotations by participants using annotation device \cite{sharma_dataset_2019, xue_rcea-360vr_2021, 8105870} and self-reflection \cite{10.1145/3334480.3383019}. Few other works have used the stimulation task as ground truth; for example, in the WESAD dataset \cite{schmidt_introducing_2018}, physiological data collected during the Trier Social Stress Test (TSST) is labeled as stress. 

\textbf{Physiological Signals and Derived Features}: The existing literature on emotion recognition explores the complex interplay between physiological signals and human emotions. Researchers have leveraged various physiological signals and their derived features, such as heart rate (HR), heart rate variability (HRV) measured using electrocardiogram (ECG), photoplethysmography (PPG) or blood volume pulse (BVP) signals, skin conductance measured using electrodermal activity (EDA), also known as galvanic skin response or resistance (GSR), skin temperature (SKT), and muscle activity measured using electromyogram (EMG) for quantifying changes in emotional arousal \cite{koelstra_deap_2012, abadi_decaf_2015, behnke_psychophysiology_2022, kutt_biraffe2_2022, saganowski_emognition_2022, park_k-emocon_2020}. Further movement data was collected using an accelerometer (ACC) or gyroscope (GYR) to identify changes in emotion. 

Besides lab-based studies, fields with constraint settings such as workspace \cite{koldijk_swell_2014, hosseini_multimodal_2022} are also explored in previous work for collecting emotion data. Daily life or field settings are also investigated but are limited in number due to the lack of control \cite{mundnich_tiles-2018_2020, yfantidou_lifesnaps_2022, kang_k-emophone_2023}. The current approaches to emotion data collection have the subsequent constraints: (1) they depend on self-reported data, and (2) they offer snapshots of specific moments rather than providing continuous monitoring and often do not value context information. Moreover, data collection practices often lack the ability to either collect or utilize the human-specific context in data collection and algorithmic designing. This study aims to systematically understand the role of participant subjectiveness and how it impacts the physiological emotion data.

\begin{figure}[ht]
    \centering
    \includegraphics[width=1\linewidth]{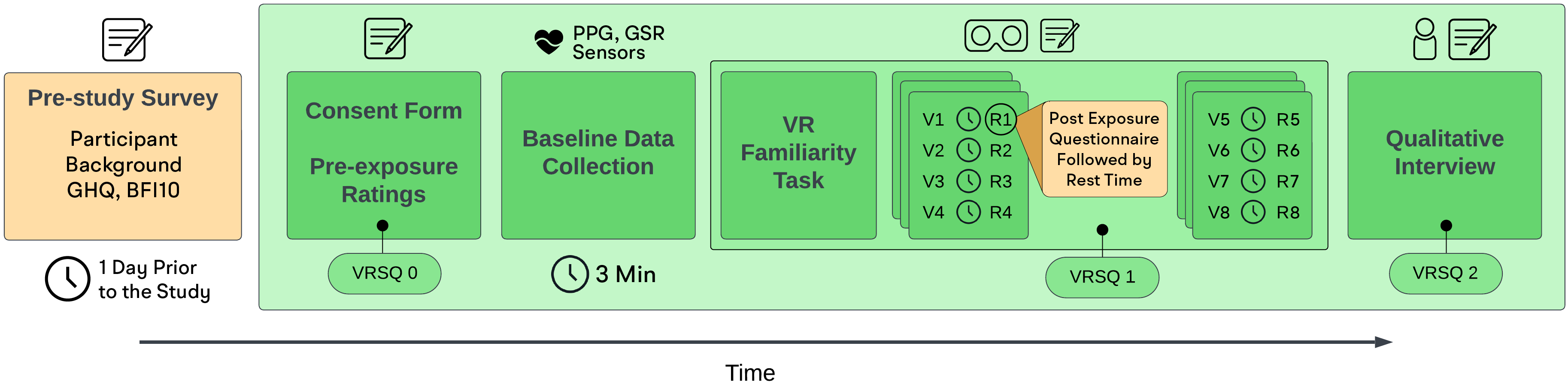}
    \caption{The figure depicts the data collection procedure used in this paper in their execution order. In the diagram, the following abbreviations correspond to the respective components: GHQ (General Health Questionnaire), BFI10 (Big Five Inventory-10), VRSQ (Virtual Reality Sickness Questionnaire), PPG (Photoplethysmography), EDA (Galvanic Skin Response), V (video or Stimulus), and R (Rest). The Pre and Post-Exposure Ratings include the Positive-Negative Affect Scale (PANAS) and the SAM Scales.}
    \label{fig:methods}
\end{figure}

\section{Methodology}

To investigate our RQs, we conducted the data collection experiment in virtual reality using VR 360{$^\circ$} short videos. The Institute Review Board (IRB) approved the study for ethical considerations.  At the end of the study, participants received merchandise goodies worth \$4 as a token of participation. The following sections describe our methods.


\subsection{Participant Selection}\label{sec:participants}

Our study comprised 37 healthy participants (21 males, 15 females, 1 undisclosed) aged 18-33 (M=23.1, SD=4.02). Recruitment was conducted through institute-wide email calls and promotions within social media circles. Exclusion criteria encompassed individuals with experience or a history of heart issues, heart arrhythmia, high blood pressure, medical conditions affecting equilibrium, visual or auditory impairments, neurological ailments, cognitive challenges, psychological issues, or diagnosed depression \cite{tabbaa_vreed_2022}. 
Additionally, participants with low proficiency in the English language were not included in the study to avoid the impact of language. Our participant demographic and data summary from our pre-study survey are presented in table \ref{tab:participant_table}. 

\begin{table*}[htbp]
\small
    \centering
    \begin{tabular}{ll}
        \toprule
        \textbf{Category} & \textbf{Details and Count} \\
        \midrule
        \textbf{Total Participants} & \textbf{37} \\
        \midrule
        \textbf{Gender} & Female \textbf{(15)}, Male \textbf{(21)}, Prefer not to say \textbf{(1)} \\
        \midrule
        \textbf{Playlist} & Playlist 1 \textbf{(6M, 4F)}, Playlist 2 \textbf{(4M, 4F)}, \\
        & Playlist 3 \textbf{(5M, 4F)}, Playlist 4 \textbf{(6M, 4F)} \\
        \midrule
        \textbf{Age} & Range \textbf{18-33}, Mean age \textbf{23.1}, Standard deviation \textbf{4.02} \\
        \midrule
        \textbf{Education} & Senior High School \textbf{(4)}, Bachelor's Degree \textbf{(24)}, Master's Degree \textbf{(8)}, Doctorate \textbf{(3)} \\
        \midrule
        \textbf{Awareness about VR} & Yes \textbf{(27)}, No \textbf{(10)} \\
        \midrule
        \textbf{Usage of VR} & Never used before \textbf{(17)}, Rarely \textbf{(14)}, Sometimes \textbf{(4)}, Often \textbf{(1)}, Very often \textbf{(1)} \\
        \midrule
        \textbf{Average Screen Time} & Less than 2 hours \textbf{(2)}, 2-4 hours \textbf{(5)}, 4-6 hours \textbf{(5)}, 6-8 hours \textbf{(8)},\\
            &8-10 hours \textbf{(11)}, More than 10 hours \textbf{(6)} \\
        \midrule
        \textbf{General Health Assessment} & Healthy \textbf{(20)}, Distressed \textbf{(17)} \\
        \midrule
        \textbf{Personality Characteristics} & 
              Agreeableness -     Low \textbf{(3)}, High \textbf{(34)} \\
            & Extraversion -      Low \textbf{(5)}, High \textbf{(32)} \\
            & Conscientiousness - Low \textbf{(9)}, High \textbf{(28)} \\
            & Neuroticism -       Low \textbf{(13)}, High \textbf{(24)} \\
            & Openness -          Low \textbf{(5)}, High \textbf{(32)} \\
        \bottomrule
    \end{tabular}
    \caption{Demographic information of the study participants, refer section \ref{datacollection} on experiment procedure for more information on Playlist, General Health Assessment and Personality Characteristics}
    \label{tab:participant_table}
\end{table*}

\begin{figure*}[htbp]
    \centering

    \includegraphics[width=1\textwidth]{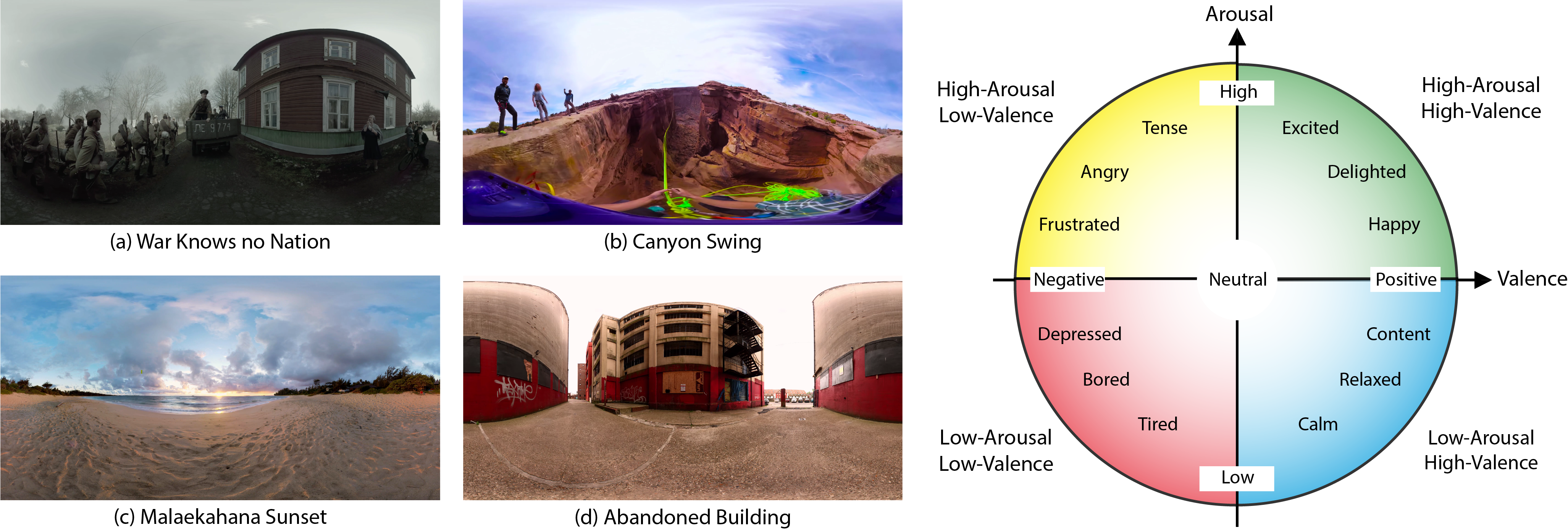}

    \caption{The figure shows stills taken from the \(360^{\circ}\) videos capturing different environments shown to the participants as a part of the experiment methodology. The stills capture different valence-arousal combinations such as (a) Low-Valence-High-Arousal (LVHA), (b) High-Valence-High-Arousal (HVHA), (c) High-Valence-low-arousal (HVLA) and (d) Low-Valence-Low-Arousal (LVLA), the database is publically available \cite{li2017public}.}
    \label{fig:samplevideo}
\end{figure*}

\subsection{Stimulus Selection and Data Collection}\label{datacollection}

\begin{figure}[htbp]
    \centering

    \begin{subfigure}[b]{0.5\textwidth}
        \centering
        \includegraphics[width=\textwidth]{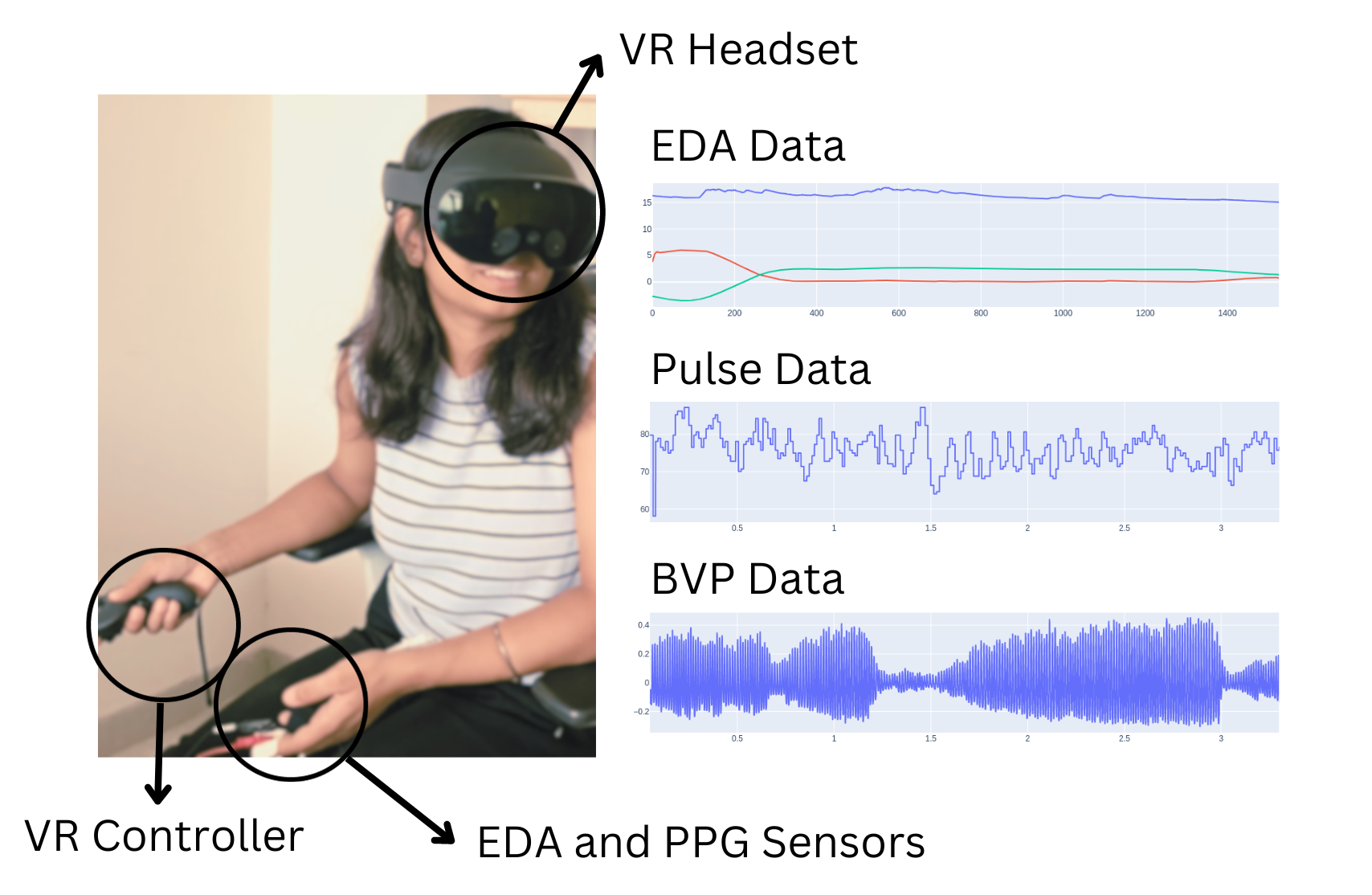}
        \caption{The experiment setup and illustration of collected physiological signal data}
        \label{fig:subfigure1}
    \end{subfigure}
    \hfill
    \begin{subfigure}[b]{0.45\textwidth}
        \centering
        \includegraphics[width=\textwidth]{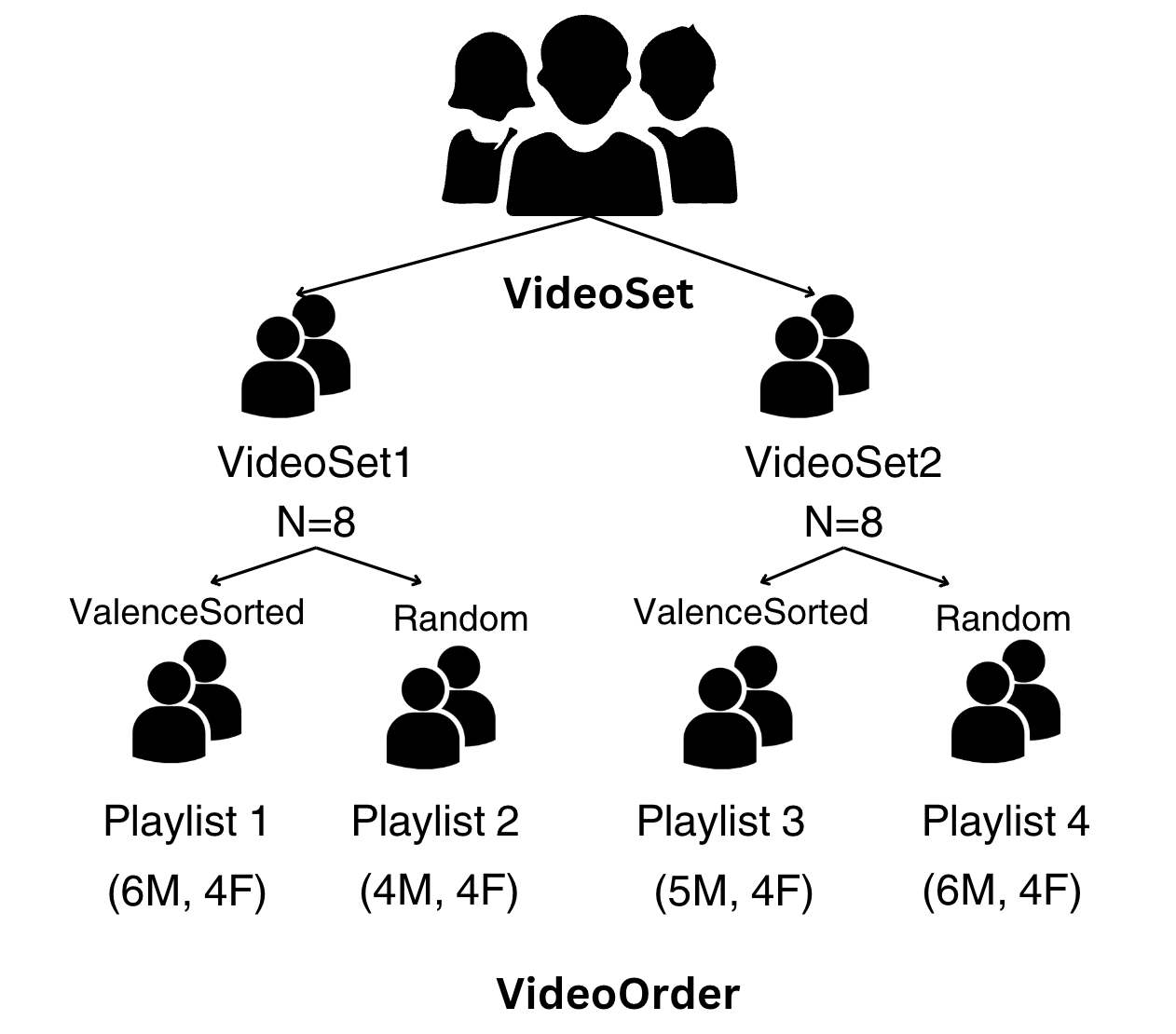}
        \caption{Between subject Methodology}
        \label{fig:subfigure2}
    \end{subfigure}

    \caption{Study Design}
    \label{fig:study_design_subfigures}
\end{figure}

We employed a between-subjects study design (Figure \ref{fig:subfigure2}) with two independent variables. The first, \emph{VideoSet}, refers to the two sets of (N=8) VR 360{$^\circ$} videos presented to participants. The second variable, \emph{VideoOrder}, pertains to the sequence in which a video set is presented to participants. 
For this experiment, a total of 16 videos were utilized. The videos were selected from a publicly available 360{$^\circ$} VR dataset \cite{li2017public}, also utilized in prior work. To curate the video subset, we applied a heuristics protocol, selecting four videos from each category of the circumplex model \cite{russell_affect_1989}. The heuristic involved choosing videos with maximum distance from the origin to enhance coverage and diversity within the subset. The 16 videos were then divided into two subgroups of N=8 videos each, considering factors such as pilot feedback, total experiment time, and VR exposure to prevent participant fatigue or motion sickness. 
The subgroup selection involved arranging videos based on their valence ratings (represents the degree of pleasantness or unpleasantness associated with an emotional state). This was done to determine if a specific sequence of emotions (from unpleasant to highly pleasant) impacts the overall emotional response of the participant, compared to a random presentation of stimuli that is independent of the valence rating. We chose valence-based ordering because it directly reflects the emotional response triggered by the stimulus and is easier for participants to identify and distinguish. In contrast, arousal is more subjective and harder for participants to differentiate \cite{kensinger2006processing, barrett2006valence, barrett1999structure}. To simplify the experiment for participants, we prioritized valence-based ordering.
Subsequently, alternate videos were paired from each quadrant to create two VideoSet. This step aimed to establish playlist normalization, ensuring a balanced experimental setting. The selected videos were downloaded from database\footnote{https://stanfordvr.com/360-video-database/} using youtube-dl\footnote{https://github.com/ytdl-org/youtube-dl} tool in the equirectangular panoramic format with a resolution of 3840 x 2160 pixels.  All the videos were edited to fit within a 3-minute timeframe, as per prior work \cite{tabbaa_vreed_2022}. To investigate the impact of video order on emotional responses, we organized the videos into two distinct orders: Valence Sorted Order, where videos were arranged based on valence ratings within a VideoSet, and Random Order, where videos were arranged randomly, irrespective of their ratings. After applying these orders, we created four playlists. The sorting technique employed for the study involved arranging videos from low negative valence to high positive valence, facilitating the determination of the emotional properties as positive or negative. The four playlists (see Table \ref{tab:playlist_table}) are as follows- \emph{Playlist1: VideoSet1 - Random Order},  \emph{Playlist2: VideoSet1 - Valence Sorted Order}, \emph{Playlist3: VideoSet2 - Random Order}, and \emph{Playlist4: VideoSet2 - Valence Sorted Order}. Participants were allocated playlists in a gender-balanced manner through random assignment (see Figure \ref{fig:subfigure2}). The videos (see Figure \ref{fig:samplevideo}) were presented to participants in a custom Virtual environment with separate scenes using Unity Engine. 

\textbf{Experiment Procedure}: Our experiment, as depicted in Figure \ref{fig:methods}, lasted approximately 1 hour and 15 minutes. Prior to data collection, an email was rolled out inviting individuals to participate in the study. Individuals who opted in for the study received an email confirming their availability. Twenty-four hours before the experiment, participants were given a pre-study survey to collect crucial participant information. Since the nature of participation was opt-in, this was considered implicit consent for the pre-study questionnaire. This questionnaire covered - \textit{Background Information}: Questions regarding gender, age, and familiarity with VR technology on a 5-point Likert scale. Open-ended questions were included to understand any prior exposure to VR. \textit{General Health Assessment}: To assess psychological well-being over the past week, we utilized the twelve-item General Health questionnaire (GHQ-12) \cite{qin2018general}. This non-medical screening tool helped gauge the emotional state of otherwise healthy participants with no diagnosed mental health conditions in the week prior to data collection. \textit{Personality Assessment}: Participants' personality types were assessed using the Big Five Inventory-10 (BFI-10) \cite{rammstedt2007measuring}. The inclusion of GHQ-12 and BFI-10 in the pre-study questionnaire was influenced by previous data collection research \cite{subramanian_ascertain_2018, miranda-correa_amigos_2021, romo2024eeg}. Other participants' parameter characteristics, like average screen time, were collected to understand participants' consumption of digital media and their technology awareness. 

On the data collection day, participants were briefed about the study without revealing its objectives. Subsequently, they read the privacy policy and the associated risks and signed the consent form. 
Following the briefing, a consent form was presented to the participants. Consenting participants were directed to complete a \textit{pre-exposure form}. This form included questions about current emotional states using standard PANAS \cite{watson_development_1988}, SAM \cite{bradley1994measuring} scales, and Virtual Reality Sickness Questionnaire (VRSQ) \cite{kim2018virtual} to assess initial fatigue and motion sickness symptoms before their VR exposure. The PANAS scale gathered positive and negative affect readings on a 5-point scale, covering ten positive and ten negative emotions \cite{watson_development_1988}. The SAM scale was utilized for dimensional ratings of Arousal (refers to the intensity associated with emotion), Valence (positivity to the negativity of emotion), and Dominance (degree of control over emotion). Participants were encouraged to seek clarification from the experimenter if they needed assistance understanding the form. 
After completing the pre-exposure form, Photoplethysmography (PPG) and Electrodermal Activity (EDA) sensors were attached to the participant's non-dominant hand fingers. Subsequently, 3 minutes of baseline data was collected, during which participants were instructed to sit and relax. Studies have shown the benefits of a shorter baseline collection period \cite{https://doi.org/10.1111/j.1469-8986.1992.tb02052.x}. Consequently, we adopted a 3-minute period for baseline data collection, aligning it with our stimulus duration of around 3 minutes to ensure a balanced amount of physiological data collected. The choice of PPG and EDA sensors was based on their widespread use in emotion recognition.
Within the VR environment, participants initially sat in a waiting room to acclimate to the technology, exploring the surroundings by looking around. Following this, participants transitioned to familiarizing themselves with the VR controller. Engaging in a simple game, they used the controller to pick up and throw a ball into a box within the VR room. After completing the VR familiarity task, participants were instructed to choose the assigned playlist and commence watching the videos in that playlist. Following each video, participants completed the \textit{post-exposure form} (same as pre-exposure form) to document their emotions during the viewing. Once the form was completed, participants rested before proceeding to the following video.

\begin{table*}[htbp]
\small
    \centering
    \begin{tabular}{ll}
        \toprule
        \textbf{Playlist Number} & \textbf{Reference Video Number: Name of Video (In Order)} \\
        \midrule
        Playlist 1 & P1V1: The Displaced, P1V2: Happyland 360, P1V3: Jailbreak 360, \\
                   & P1V4: War Knows No Nation, P1V5: Canyon Swing, P1V6: Redwoods Walk Among Giants, \\
                   & P1V7: Speed Flying, P1V8: Instant Caribbean Vacation \\
        \midrule
        Playlist 2 & P2V1: The Nepal Earthquake Aftermath, P2V2: Zombie Apocalypse Horror,\\
                   & P2V3: Abandoned building, P2V4: Kidnapped, P2V5: Mega Coaster, \\
                   & P2V6: Malaekahana Sunrise, P2V7: Puppies host SourceFed for a day, \\
                   & P2V8: Great Ocean Road \\
        \midrule
        Playlist 3 & P3V1: War Knows No Nation, P3V2: Redwoods Walk Among Giants, P3V3: Happyland 360, \\
                   & P3V4: Speed Flying, P3V5: Instant Caribbean Vacation, P3V6: Jailbreak 360, \\
                   & P3V7: The Displaced, P3V8: Canyon Swing \\
        \midrule
        Playlist 4 & P4V1: Kidnapped, P4V2: Malaekahana Sunrise, P4V3: Zombie Apocalypse Horror, \\
                   & P4V4: Puppies host SourceFed for a day, P4V5: Great Ocean Road, \\
                   & P4V6: Abandoned building, P4V7: The Nepal Earthquake Aftermath, P4V8: Mega Coaster \\
        \bottomrule
    \end{tabular}
    \caption{List of videos in each playlist arranged in the order as shown to participants}
    \label{tab:playlist_table}
\end{table*}





\subsection{Experiment Setup}
The study was conducted in the institute's research lab. Our experiment setup is depicted in figure \ref{fig:subfigure1}.

\textbf{VR and Questionnaire Setup}: Meta Quest Pro was used to show the VEs. The headset has 2 x LCD panels with 1800 x 1920 pixels per eye, a refresh rate of 90Hz, 106º Horizontal × 96º Vertical Field of view. Additionally, it incorporates eye relief adjustment, lens spacing features, and Spatial audio support. ASUS TUF laptop was used to run the project with 11th Gen Intel core and NVIDIA GeForce RTX 3050. The OpenXR plugin integrated the Meta Quest pro headset with the Unity application. OpenXR plugin helped us with hand gestures and controls for interacting with the Application User Interface. An iPad Pro tablet was used to complete the pre and post-exposure forms.

\textbf{Physiological Measure Apparatus}: A 4-channel Biopac MP36 \cite{biopac} was used for continuous collection of PPG (SS4LA Hardware module) and EDA (SS57LA Hardware module) data using separate channels. BSL 4 software from biopac was used for data acquisition. The EDA sensor was attached to the index and middle fingers \cite{tabbaa_vreed_2022} of the participants' non-dominant hand, utilizing EL507 Electrodes for collecting users' skin electrical conductance. The PPG sensor was attached to the participant's non-dominant hand ring finger. Before attaching the sensors, Isotonic Gel was applied, and the electrodes were firmly fastened to ensure minimal noise in the collected data. 

\subsection{Interviews}



We conducted a concluding semi-structured interview with the 37 participants post-exposure to the 360{$^\circ$} VR stimulus. The interviews were conducted in English and Hindi as required after obtaining consent to record audio. 
During interviews, our objective was to grasp participants' emotional responses, comprehend the reasoning behind these feelings, gain insights into their understanding of scales employed for emotion annotations, and assess the impact of the experimental design and setup on their emotions. 
The interviews began with an opening question, \emph{"How was your overall experience?"} We then assessed participants' emotional engagement to study RQ1 by showing brief segments of each video on a computer screen to aid recollection. Following this, we explored the dominant emotion in each video, prompting participants to articulate their emotional responses and reasons behind annotation by referring to their subjective ratings.
To investigate RQ2, we asked questions such as: \emph{"How easy was it for you to transition emotions between videos? Did breaks between videos aid emotional transitions? Did preceding videos influence your emotions towards subsequent ones? What prevented you from feeling the expected emotion tied to the stimuli? Did you face any challenges filling out the pre- and post-exposure questionnaires?"}
Subsequently, to understand the impact of VR 360{$^\circ$} videos as an elicitation medium, we focused on various aspects of viewing experiences in VR. This encompassed elements such as camera angles, VR content stitching quality, different perspectives, overall video quality, and impact of real-life camera footage in comparison to computer-generated imagery (CGI) (Real-life footage such as "Happyland," filmed in a real-life slum area in the Philippines whereas computer-generated videos featured scenes created using visual effects and computer-generated-imagery, such as "War Knows No Nation," which used CGI to depict soldiers from various countries worldwide). We also delved into the auditory experience in VR, considering background music (such as calm music, energetic music, lyrical songs, and other types of background audio like birds chirping, sea waves sound, etc., as per the stimulus content), audio clarity, and the spatial orientation of audio cues. The linguistic experience of stimuli was examined, including factors like the language of narration and the presence or absence of subtitles. Additionally, we considered the influence of the lab setting and participants' awareness of it on their experiences. Our interviews followed a semi-structured approach, allowing us to adapt the questions based on participants' feedback.

\section{Data Analysis}
For analyzing our interview data, we first transcribed the recorded audio data into English using the Google Cloud speech-to-text API \textcolor{blue}{\footnote{https://cloud.google.com/speech-to-text}} as an initial draft. This draft was then checked for transcription and spelling errors by the first and second authors, who manually matched the text to the audio recordings to correct any mistakes (such as if heavily accented words were identified correctly or if there were any transcription mistakes in general) made by the Google API. We also manually added speaker identifiers (participant and interviewer) and identified text segments corresponding to particular videos. We have adopted realist epistemology \cite{papineau1985realism} for analyzing our qualitative data, wherein we have reported the emotional experiences and perceptions of the participants while experiencing the stimulus and labeling the emotions.
Subsequently, we employed inductive thematic analysis \cite{braun2006using} on the refined transcripts from the previous steps. In the initial phase, the first two authors performed open coding on the interview transcripts line-by-line. Codes were reviewed and aligned to ensure consistency, with all authors collectively involved in the analysis process. Sample codes from our open coding included \emph{"Novelty of VR," "connection with past," "future aspiration,"} and \emph{"expectations from VR"}, etc. These open codes were then clustered based on emerging patterns over multiple iterations to derive axial codes, such as \emph{"drawbacks of VR," "participant's perspective,"} and \emph{"relationship with stimuli"}, etc. Finally, we used selective coding to refine the axial codes and identify the final themes that guided the structuring of our findings. Throughout this process, the research team utilized tools like Miro-board \footnote{https://miro.com/} and Google Sheets for conceptualizing and theme construction. We performed statistical tests on both EDA and PPG data for our quantitative data analysis. For EDA, we extracted the mean EDA values. For PPG, we extracted HRV RMSSD (Root Mean Square of Successive Differences) from the signal data collected during baseline and stimulus periods. These features were selected as indicators of parasympathetic nervous system activity \cite{schmidt_introducing_2018}.
Initially, we tested the data for normality using Q-Q plots and the Shapiro-Wilk test. Since the data did not follow a normal distribution, we applied a two-tailed Wilcoxon rank-sum test to compare the distributions of two independent groups for our analysis and the Kruskal-Wallis test for more than two groups. 

\section{Study Limitations}
A significant constraint of our study is the generalizability of our findings because of our participants' cultural background and technology awareness. Despite employing diverse recruitment methods to ensure varied participation, our current demographic primarily comprises tech-savvy individuals aged 18-33 with a higher level of education. It is crucial to acknowledge that our findings may be influenced by participants' age, educational backgrounds, technology experience, and self-emotional understanding. While these limitations exist, it is essential to recognize that emotion elicitation is inherently challenging and strongly influenced by personalized factors. Our study underscores the significance of understanding and collaboratively involving human participants in emotion data work.
Additionally, we stress that our study, conducted in controlled lab settings, may not readily extend to data collection in natural, real-world environments. However, despite taking place in controlled lab environments using passive virtual reality stimuli, we believe some of our findings apply to the broader paradigm of physiological emotion data collection studies, while some remain specific to VR-based elicitation. We recognize that a more extensive and diverse sample encompassing various perspectives would offer deeper insights, and we plan to explore this aspect in our future work.

\begin{table*}[htbp]
\small
    \centering
    \begin{tabular}{ll}
        \toprule
        \textbf{Category} & \textbf{Key Insights} \\
        \midrule
        \textbf{Participant perspective} 
            & - Self perception \\
            & - Prior knowledge and experiences with stimulus\\
            & - Logically reasonable and emotionally relevant stimulus \\
            & - Motivation towards experiment\\
        \midrule
        \textbf{Experiment Design} 
            & - Order and length of stimulus \\
            & - Exposure duration according to the emotion of stimulus \\
            & - Interval between different stimuli \\
            & - Choice of self-annotation method \\
        \midrule
        \textbf{Virtual Reality}
            & - High immersion and presence \\
            & - Need for Interaction within stimulus \\
            & - Presence of human-like elements \\
            & - Quality of video and audio elements \\
        \bottomrule
    \end{tabular}
    \caption{Summary of the Findings}
    \label{tab:finding}
\end{table*}

\section{Findings}

In this section, we present the implications and influence of participant perception, experiment design, and experiment setup choices on a participant's emotional response and annotation. The summary of our findings is presented in Table\ref{tab:finding}.  

\subsection{Unveiling the Influence of Participant Perception}
\subsubsection{The Influence of Participant's Self-perception}

Our interviews revealed that participants \textit{self-perception} played an important role in how they responded to a stimulus. On various occasions, participants pointed out that their typical behaviour patterns and personality traits in daily life have shaped their emotional response to a stimulus shown in the experiment. We found that participants who self-perceived themselves as emotionally stable or in control of their emotions were less responsive towards the stimulus than participants who did not have a strong self-perception of being emotionally stable. A participant shared:
\begin{quote}
"Not much change in emotion I would say. Not sure but I feel I have control over my emotions to an extent, I think so I am normally neutral in most of the situations." \textbf{[P39, Playlist3]}
\end{quote}
Further, it was also observed that participants with strong self-perceptions of being emotionally stable consciously chose not to respond to a negative stimulus, as explained by a participant \textbf{[P33, Playlist 3}] \textit{"I know this is sad, but I won't cry over it"}. 
Similarly, individuals who identified as nature or pet enthusiasts displayed heightened emotional responses to stimuli involving natural settings or animals. For example, those who preferred beaches appeared more serene when viewing videos of beach scenes. In contrast, participants who were neutral towards puppies demonstrated a subdued reaction, whereas those with a fondness for puppies exhibited a more pronounced emotional response. Overall, participants' self-perceptions have shaped their responses to a stimulus and their annotations irrespective of the targeted stimulus emotion.

\subsubsection{The Influence of Perception towards the Content of Stimulus}

All participants mentioned that their viewpoint toward the content of the stimulus was a major reason behind eliciting a response.  
Past experiences were frequently linked to the content displayed in the stimulus, resulting in increased engagement among participants. Recalling familiar life scenarios made participants feel more emotionally connected to the stimuli. For example, participants mentioned instances such as trekking to locations resembling the stimulus and experiencing earthquakes in the past. One of the participants explained:
\begin{quote}
"Like I was sad...because one of my friend is in army, so I felt like he is going on war...yeah so one part where the woman is outside the train and the army was like going- from the window she was waving" \textbf{[P29, Playlist1]}
\end{quote}
Similarly, prior knowledge about the content shown in the stimulus also contributed to the emotional response. Participants reported that prior knowledge has impacted their emotional engagement; for instance, prior knowledge about World War II and that it had already happened in the past has led to emotional detachment towards the content. A participant [\textbf{P14, Playlist 2}] explained that \textit{"No, because I am aware that it happened [World War 2], so it did not feel like it's happening right now".} According to our participants, cultural background also had a subtle impact on how much they related to content. A participant felt a stronger connection to the people of Nepal, as depicted in \textit{The Nepal Earthquake Aftermath} because they live in a geographic region closer to Nepal, while a sense of detachment was reported by participants while watching the stimulus with the story of people in Ukraine. Personal beliefs are another factor, as mentioned in our interviews, that played a role in how participants perceived and annotated a stimulus. For instance, people who value social causes or believe in inequality of resources were influenced more by stimuli like \textit{Happyland 360} than those who do not have strong opinions about poverty. 
Aspirations were another common factor often associated with the stimulus. In case of stimulus wherein adventure sports or new countries were shown \textit{Canyon Swing, Speed Flying, Great Ocean Road}, participants reported being excited as they aspire to indulge in the adventure sport or travel to that country. Next, participants revealed that cognitive processes like logical reasoning and attention toward stimulus have shaped their response to a stimulus. Participant [\textbf{P23, Playlist 3}] explained, \textit{"One thing that I could not understand logically was that I felt that the rope is so far, how come I am hanging in the middle now. I didn't understand that thing, but the rest was fine"}. We also observed logical reasoning, like an earthquake is a negative situation, and thus, the annotations should also be negative in our interviews.

\subsubsection{The Influence of Perception towards Experiment}

Through our interactions and observations, we found that participants' perception of the data collection experiment also shaped their emotional response to a stimulus. We observed in our interviews that some participants were more
interested in experiencing VR technology than others, and thus, their emotional responses were mostly excitement or boredom, depending on the stimulus. For instance, [\textbf{P3, Playlist 3}] responded to our question on \textit{How was your overall experience of the study? - It was nice. First, the videos were a little boring, but it had some interesting ones}. Another participant shared her excitement:
\begin{quote}
"The overall experience was pretty cool, the VR headset seemed perfectly fine and it was really immersive... I was on the ship, in the VR headset, it was very aesthetic and actually it made me more enthusiastic and excited to see virtual objects that seem to be real but aren’t,  so it was really amazing" \textbf{[P12, Playlist1]}
\end{quote}

Participants mentioned getting confused in some particular stimulus that didn't provide any context on what was to be shown in the stimulus for example, stimulus like \textit{Abandoned Building} just started with scenes from an abandoned area with no instructions to participants on what they are expected to do or feel. This lack of clarity on what is expected within a stimulus has created confusion and also made participants ponder on what was the targeted emotion of the given stimulus. They reported that this confusion has propagated to their self-reports as they could not understand their emotions clearly. Such a situation is often raised in videos where participants were expected to sit calmly and listen to music, while participants were expecting some cognitive load or some storytelling due to their experiences in the previous stimulus. [\textbf{P31, Playlist 1}] stated on that \textit{"I was confused in this one because nothing was happening, yeah. But then I realized that nothing was gonna happen... I was just looking around the buildings and stuff"} after watching \textit{Abandoned building}. The time of the experiment and mood before the experiment were other defining factors for the perception of the experiment. A participant reported feeling tired and less attentive to the stimulus due to their workload on that particular day and time before the experiment, which made them less interested in the experiment.

\subsubsection{The Influence of Stimulus on Participant's Self-report and Physiological Signals}

To study whether there was a change in annotations as per stimulus, we performed statistical analysis on arousal and valence scores. Our test for \textbf{change in arousal in high vs. low arousal stimulus}, showed \textbf{significant difference} (Test Statistic = 2.691, p-value = 0.007**), and for \textbf{change in valence in high vs. low valence stimulus} showed \textbf{significant difference} (Test Statistic = 7.697, p-value = 0.000**), suggesting the overall annotations have changed according to the change in stimulus.
Furthermore, to study the impact of change in stimulus on the physiological signal data of our participants, we have performed statistical tests on our participant's data.
On testing for changes in HRV RMSSD and EDA mean across different stimulus categories, we found no significant differences. Specifically, we observed no significant difference between HRV RMSSD and EDA mean values of High Valence vs. Low Valence stimuli and High Arousal vs. Low Arousal stimuli using a two-tailed Wilcoxon rank sum test. Additionally, when comparing the circumplex categories (LVLA, LVHA, HVLA, HVHA) for stimuli, no significant difference was detected in physiological signal values using the Kruskal-Wallis test. Suggesting that there were no significant changes in the HRV RMSSD and EDA mean features of the participant's data when stimulus categories were changed.

\begin{table*}[h]
\small
\centering
\begin{tabular}{lllll}
\toprule
\textbf{Category} & \textbf{Measure} & \textbf{Significance} & \textbf{p-value} & \textbf{Test-statistics} \\
\midrule
VideoSet1 vs. VideoSet2 & Arousal \textbf{(H14A)} & \textbf{S} & 0.000** & 6.57 \\
 & Valence \textbf{(H14B)} & NS & 0.372 & -0.893 \\
 & EDA Mean \textbf{(H14E)} & \textbf{S} & 0.003** & 3.007 \\
 & HRV RMSSD \textbf{(H14F)} & NS & 0.945 & 0.069 \\
\midrule
Playlist 1 vs. Playlist 3 & Arousal \textbf{(H15A}) & NS & 0.26 & -1.126 \\
 & Valence \textbf{(H15B)} & NS & 0.943 & 0.071 \\
 & EDA Mean \textbf{(H15E)} & NS & 0.358 & 0.919 \\
 & HRV RMSSD \textbf{(H15F)} & NS & 0.738 & -0.334 \\
\midrule
Playlist 2 vs. Playlist 4 & Arousal \textbf{(H16A)} & NS & 0.359 & 0.917 \\
 & Valence \textbf{(H16B)} & NS & 0.122 & -1.547 \\
 & EDA Mean \textbf{(H16E)}  & NS & 0.5 & 0.674 \\
 & HRV RMSSD \textbf{(H16F)} & \textbf{S} & 0.021* & 2.306 \\
\bottomrule
\end{tabular}
\caption{Summary of results of statistical tests for analyzing the impact of Video Set and order on both Subjective measures and Physiological data. In the table NS: Non-significant, S: Significant, **: less than alpha 0.01, *: less than alpha 0.05}
\label{tab:testtable}
\end{table*}

\subsubsection{The Influence of Participant's Characteristics on Physiological Signals}

For analyzing GHQ data, we calculated the GHQ score by summing the responses to the 12 questions, resulting in a total score ranging from 0 to 36. A lower score indicates better mental health and less psychological distress, while a higher score suggests elevated levels of psychological distress over the past week. In our data, 17 out of 37 participants have a high GHQ score, which suggests that these participants have elevated psychological distress in the past week (in real life outside our study environment). To test the impact of GHQ scores on physiological data (collected during baseline+stimulus), we conducted a two-tailed Wilcoxon rank-sum test to compare the EDA mean and HRV RMSSD values between participants with High GHQ scores vs. Low GHQ scores. The results showed no significant differences in EDA mean or HRV RMSSD between the two groups. We also conducted individual tests on the extracted values for the baseline and stimulus data separately but found no significant differences in these subsets either. This suggested that GHQ scores don't have any impact on the physiological data on our subset of participants. We conducted a series of tests to analyze the impact of different personality characteristics on physiological data. First, we calculated personality scores for each category: Agreeableness, Extraversion, Conscientiousness, Neuroticism, and Openness (more details in Table \ref{tab:participant_table}). We then applied a two-tailed Wilcoxon test on the combined data (baseline+stimulus), baseline-only, and stimulus-only data.
For participants with High vs. Low Agreeableness, no significant difference in physiological data was found. The same result was observed for Extraversion. However, for Conscientiousness, we found a \textbf{significant difference} in HRV RMSSD data (baseline+stimulus) between High and Low Conscientiousness participants (Test Statistic = -3.263, p-value = 0.001**) but no significant difference in EDA mean data.
For Neuroticism, a \textbf{significant difference} was observed in EDA mean data (baseline+stimulus) between High and Low Neuroticism participants (Test Statistic = 3.818, p-value = 0.000**). Additionally, a \textbf{significant difference} in HRV RMSSD was found in the stimulus-only data for Neuroticism (Test Statistic = -2.120, p-value = 0.034*). Lastly, for Openness, we observed a \textbf{significant difference} in HRV RMSSD data (stimulus+baseline) between High and Low Openness participants (Test Statistic = -3.865, p-value=0.000*), while the EDA mean was non-significant.

\subsection{Deciphering the Experiment Design's Impact and Significance}
\subsubsection{Order, Length and Choice of Stimulus matters}
In our interviews with participants, we noticed that experiment design played an important role in emotion data collection. Design choices like the order in which the stimulus would be shown to participants and the length of the stimulus mattered to our participants. In our interviews, participants reported anticipating something to happen in a calming stimulus because the previous stimulus they had seen involved a lot of actions (e.g., observed in playlist 3 - P3V1 (War knowns No Nation) followed by P3V2 (Redwood Walks among Giants)). This suggested an impact of order in the form of anticipation and expectation from the content of the stimulus, impacting an otherwise relaxed emotional response. A participant stated:
\begin{quote}
"At first I was like, kind of alert, like in that question [PANAS] because I thought something was going to happen, because after the first video, I thought in the second video something is gonna happen like some jump scare or someones gonna come or something will happen. But then I realised it is a calming situation." \textbf{[P12, Playlist4]}
\end{quote}

The stimulus length was another important factor influencing the participant's emotional response. Participant [\textbf{P9, Playlist 2}] stated \textit{"Yeah, it was calm...But then, I think it went on for too long...So, this got boring"}. Instances were noted where participants expressed boredom during lengthy stimuli, dissatisfaction with insufficient exposure, and an inability to feel any emotion during short exposures. Notably, participants found it easier to connect with brief positive stimuli, whereas more time was deemed necessary for negative stimuli. One participant articulated that the stimulus duration was too brief for her to establish an emotional connection. This highlights the nuanced relationship between stimulus length and participants' emotional responses. Additionally, we performed statistical tests to analyze the impact of VideoSets and the order of stimulus (playlists) on participants' physiological data and their annotations (valence and arousal). We found a \textbf{significant difference} in the Arousal score and EDA mean of participants in VideoSet1 vs. VideoSet2. Suggesting that VideoSets have an impact on arousal scores and EDA data. As for playlists, we found a \textbf{significant difference} in HRV RMSSD for participants in Playlist 2 vs. Playlist 4. This may or may not be due to playlist order since HRV RMSSD can also be influenced by the participant pool in the playlists. More details are provided in Table \ref{tab:testtable}.

\subsubsection{Challenges in Annotating the Emotions}
Besides the order, length, and choice of stimulus, the annotation method was another crucial element in the data collection procedure. In our interviews, participants mentioned that they faced issues annotating their emotions on the SAM scale when they faced mixed emotions like fear and excitement [In Speed Flying], while PANAS was better for annotating mixed emotions. Another common comment by participants was that they found it hard to quantify their emotions in numbers and preferred interviews to express their emotions as stated by [\textbf{P9, Playlist 2}] \textit{"this is really hard to answer, yeah, you can articulate it but actually quantifying it is a very difficult task to do so, overall you can say, okay, four or five or three [referring to SAM scale], I was a bit exciting, more exciting, but how much that is really difficult"}. Most participants noticed that few emotions were part of the PANAS scale, like, \textit{hostile or guilty}, but they never felt those emotions throughout the experiment, questioning the need for including them in the questionnaire. A participant compared the two scales and explained:
\begin{quote}
"So I think partly name of the emotion [PANAS], because I was able to like correlate with words more than quantify a metric to my emotion, and like with the diagrams [SAM Scale] and all that, how am I feeling." \textbf{[P25, Playlist2]}
\end{quote}
Participants also pointed out a need for randomness in the self-report questionnaire as they were repetitive after each stimulus, giving them a chance to fill it out without much thought. Gaps or rest periods outside the VR environment between the consecutive stimuli were reported to help transition from one emotion to another, and thus, they preferred filling their annotations outside the VR environment. 

\subsection{A participant's (Virtual) Reality - Implications of Experiment Setup}
\subsubsection{Suitability of VR as an Elicitation Medium}

Through our interviews, we observed that there were certain aspects of VR as a medium that impacted the overall experience of the participants, both in a positive and a limiting manner. Many participants mentioned that they could feel being present in the setting where the stimulus was set. This heightened sense of presence not only improved participants' overall experience but also contributed to a more profound emotional engagement with the stimulus. For example, as reported by [\textbf{P2, Playlist 1}], \textit{"I just wanted to go back in time and see how people were doing things... I can't go there, but I was able to feel all that"} [referring to soldiers in war]. Participants also mentioned that when they immersed themselves in the VR content, they became somewhat unaware of their real-life surroundings and reported being more engaged with the content. A participant explained:
\begin{quote}
    "And I'm a huge roller coaster fan in general, so this was the first time I actually felt like I was in it, like I could like physiologically feel like okay I'm taking a turn- yeah my heartbeat was… and I don't know there's this one feeling, like something happening in my stomach but in a good way, like in an excited way" \textbf{[P18, Playlist4]}
\end{quote}
Many participants reported feeling the urge to interact with the stimulus, but since it was a video-based stimulus, they could not do so. Participants also reported feeling that their actions were restricted because it was a passive stimulus. Some participants mentioned that in addition to the visual stimulation, they would also want some physical sensation that compliments the presented stimulus to feel more present in the stimulus. For example, [\textbf{P27, Playlist 4}] mentioned, \textit{"I felt like I was there, I could- there was a person sitting behind as well, there was some wind, but I could not feel the wind, like I could listen to the wind" }, since they were immersed in the video but their experience got hindered due to a lack of the physical stimulation that comes with the wind brushing past.
Some participants had biases toward VR, which hindered their overall experience. As [\textbf{P3, Playlist 3}] mentioned, VR seemed to be \textit{"a little on the fake side"} to them. VR was a first-time experience for many participants, and they reported having elevated expectations from VR, which was influenced by the amount of knowledge they had about the medium. [\textbf{P17, Playlist 3}] mentioned that they were curious as to how much can one experience within a VR setting and were wondering, \textit{"somebody can be sitting in their chair, in their house, and could be feeling like they are in the woods surrounded by such huge trees"}. Some participants also mentioned the opposite, that due to their past experiences with VR, they were pretty familiar with the platform and had realistic expectations from the various stimuli presented to them. A participant explained:
\begin{quote}
    "VR usually means there is a purpose to the VR content that is created, and here I felt like it was just sort of being around nature. That was the only context, and for that I think I should be there, like I expected smell, I expected, like, the environment, the atmosphere, maybe some animal sounds, but those things weren't there [in VR stimuli]" \textbf{[P21, Playlist1]}
\end{quote}

\subsubsection{Constraints of Stimulus in VR}
Besides the overall experience that the participants had with VR as a medium for stimulation, our interview questions specific to each video being presented gave us insights as to what elements of the stimulus were responsible for the experience that the viewer had. Most participants mentioned that good video quality added to their viewing experience. In addition, we observed that many participants agreed that despite the poor video quality for certain stimuli, they were willing to ignore it if they found the stimulus engaging enough. [\textbf{P3, Playlist 3}] mentioned, \textit{"I noticed the poor quality, but I liked it anyway. Video quality was not that important for me; it was the content of the videos"}. Many participants did not like the presence of quick camera cuts and fast-paced scene changes since it did not allow them to explore their surroundings. [\textbf{P19, Playlist 3}] mentioned, \textit{"this felt like immersive because I was given time to like, explore around, [and see] what's happening"}. In addition to fast cuts, participants felt that unrealistic angles and non-human camera point-of-view (POVs) created a sense of detachment from the presented stimulus. POVs that were more human-like were appreciated. For example, in the stimuli \textit{Happyland 360}, [\textbf{P39, Playlist 3}] referred to the eye-level placement of the camera and mentioned, \textit{"I actually felt connected because most of the children- they were looking into the camera so it felt like they are looking at me, and we have a direct eye contact"}. Stimuli with video stitching issues and \textit{black holes} at the bottom of the recording hindered the participants' experience. For instance, when asked \textit{"The camera angle was such that you were the head of the person, did that create any impact?"} for the stimulus \textit{Kidnapped}, [\textbf{P4, Playlist 4}] answered, \textit{"No, because I looked down and I could see like a black hole there"}. The presence and absence of human elements within the video stimulus was also a factor, and participants reported that human interactions within the stimulus increased their sense of being present in the stimulus scenario, as explained by a participant:

\begin{quote}
    "It was more [real] because of the environment...like if I see down there is debris and there is someone else [in stimulus] also standing with me, so I could relate because he's there, I'm also there. So the environment or the other person made me feel that I was there [in the video]" \textbf{[P27, Playlist4]}
\end{quote}

Audio was also an equally important aspect that all participants focused on. For stimuli that had a narration or a human talking in a language that the participant could not understand, the viewer reported having lower engagement levels due to the linguistic barrier. Many participants mentioned that the background music and ambient sounds, if done in moderation and matched the context of the stimulus, added to their sense of being present. A participant explained:
\begin{quote}
    "It helped, it was going with the whole experience because it was a calming kind of audio so the experience is soothing and the audio is also suiting, so it makes the whole overall impact more suiting." \textbf{[P27, Playlist4]}
\end{quote}

Participants reported having a narration introducing the stimuli, and the context was very important. It was also recorded that the narration element of the stimulus should be subtle and not \textit{overdone}. Participants reported being disengaged when the narration dictated what emotion they should feel. [\textbf{P19, Playlist 3}] reported that for the stimulus \textit{Instant Caribbean Vacation}, they felt that \textit{"The narration, I think, killed it off, it felt like an advertisement. Like, when she is speaking, I am not able to feel anything"}. Many participants mentioned that unrealistic audio can make the stimulus feel like a movie, thus detaching them from it.

\section{Discussion}

Our findings highlight the significant impact of participant-specific factors and experiment decisions, including the use of VR-based stimuli, stimuli choices, and the data labeling process, on data. These aspects resonate with participant-centric data work and the collaborative technological development themes within the CSCW community. In the subsequent sections, we delve deeper into our findings and have discussed on their implications on data quality. We have then offered recommendations to CSCW and AI researchers for future data work and experiment designs.

\subsection{Challenges and Opportunities for Participant-Centric Data Collection}

Our findings have highlighted two major aspects that play a crucial role in quality emotion data collection - 1) The role of participants and their meaning-making processes, and 2) The role of data collection setup and experiment choices.
Previous studies on emotion data collection have typically treated participants as a collective with shared characteristics, often defined by criteria such as similar backgrounds (age, gender, education and nationality), health conditions (both physical and mental), language proficiency, susceptibility to motion sickness (in the case of virtual reality), and personality traits \cite{gjoreski2017monitoring, miranda-correa_amigos_2021, hosseini_multimodal_2022}. Further, the annotations are limited to self-reporting using objective scales like Likert, SAM, or some specific questionnaires, observer's rating, stimulus label, and physiological changes \cite{schmidt_introducing_2018, saganowski_emognition_2022, gjoreski2016continuous, 9779458}. However, our findings revealed that participant context encompasses more than just the experimental setup (e.g., room temperature, sensors used, stimulus choice, and stimulus order), demographics, and physical activity. It also includes finer contextual details such as self-perception, relationship with the stimulus, and attitude towards the experiment. These factors significantly modulate emotional responses, leading to varied annotations for the same stimulus. Contextual factors such as personal preferences, prior knowledge, experiences, aspirations related to the stimulus, logical reasoning, attitude towards the experiment, and expectations from the experiment are important guides to annotations. Each of these factors can decide how different one participant's annotations would be from another participant.
Personal preference towards content can lead to favorable positive annotations to an otherwise negative stimulus. Similarly, prior knowledge about a stimulus can lead to an emotional detachment or low emotional reaction to the stimulus in the present. Prior experience of what is shown in a stimulus tends to increase the relatability with a negative stimulus while it tends to reduce novelty from a high arousal positive (exciting) stimulus. The aspirations and expectations towards a stimulus can have similar effects as personal preferences since a stimulus aligning with expectations or aspirations can lead to a positive experience while a stimulus against them can cause a lack of interest among participants. Similarly, the attitude towards the experiment is another important factor that can decide how participants annotate and experience the stimulus. In our case, we found participants were interested in experiencing VR and thus were excited by the initial stimulus irrespective of the content because they were experiencing VR, not the stimulus and its content. Logical reasoning also tends to impact the annotations and often leads to human biases, such as not emotionally engaging in the stimulus. The participant-specific factor
can thus lead to annotations that are biased by elevated or uninteresting experiences and may not be true representatives of emotions felt by participants and thus can reduce the quality of collected datasets. These contextual factors can also lead to an imbalance in the emotion data collected (For instance, if all participants felt only positive emotions because the setup was exciting or if they showed a lack of interest in negative stimuli because it was based on a past event like World War II), suggesting that equal distribution of stimuli according to the targeted emotions doesn't exactly lead to the collection of balanced amount of data for all emotions.  
Our statistical analysis also revealed that participants were able to annotate their emotions according to changing stimulus categories, while their physiological responses did not show significant changes. This suggests that participants rely more on their self-understanding and experiences for interpreting their emotions rather than on their physiological reactions, which suggests that annotations are a response to perceived emotions guided by participants' meaning-making. Our findings also aligned with Ricoeur's framework \cite{ricoeur1975philosophical}, which suggests that participants' annotations are shaped by their self-narratives and cognitive reflections rather than just physiological changes. This perspective highlights the importance of considering participants' mental and cognitive abilities in interpreting emotions instead of relying solely on self-reports, physiological data, and contextual information like activity data and personality traits \cite{miranda-correa_amigos_2021, subramanian_ascertain_2018, gjoreski2017monitoring}. Our results also align with the appraisal theory \cite{scherer1999appraisal} and the social constructivist theory of emotion \cite{barrett2017theory}, which suggest that individuals' explanations, interpretations, past experiences, cultural background, and context play a pivotal role in the experience of emotions, regardless of accompanying physiological changes \cite{aronson2005interpersonal}. 

Furthermore, we have also identified the positive and limiting effects that the choice of elicitation medium can have on the emotional response of participants. Our findings revealed that utilizing elicitation mediums like VR for emotion elicitation does have a significant impact in increasing the presence and immersion \cite{slater1993presence} and thus better elicitation. Nonetheless, we found an opportunity for further research in designing data collection experiments within CSCW communities that should consider the bias that expectations (for and against the technology) related to VR technology may introduce in emotion data collection. Moreover, our findings highlighted the importance of designing stimuli where participants can have close-to-real-life experiences with scenarios that make sense in real life, such as human-level camera angles, the presence of human elements, and active design elements. Lastly, we identified that while designing experiments for data collection studies in laboratories, careful attention should be given to the ordering of stimuli. While we did not find any statistical difference between playlists, qualitative data suggested the impact of order in some specific scenarios. Elements such as whether the stimulus has a story, includes narration, or requires participant action versus no action are crucial. These factors influence the arrangement of stimuli and the instructions to be provided to participants for a smoother emotional experience. Our results suggested that the length of stimulus is correlated with the emotion and intensity of emotions. For instance, emotions that don't require a deeper connection with a stimulus can be shorter in length, while the emotions that take a longer time to develop should have longer stimulus. We also found that annotating emotions using objective scales was challenging for our participants. In contrast, they found interviews to be an easier medium for introspection and emotional interpretation, as they were prompted and encouraged to reflect on their feelings. In the following section, we will discuss the avenues for accommodating participants' subjectiveness within data work and its implications for future work within CSCW and HCI communities.

\subsubsection{Incorporating Participant's Context}

Participant's context is crucial to how they annotate emotions. These contextual factors are the major reason behind participant's annotations. For emotion recognition models to work better, it is important that an annotation method accurately and descriptively captures the emotions a participant has felt and the context behind them. Our findings present an opportunity for the CSCW, HCI, and AI communities to develop collaborative experiment designs that acknowledge the burden placed on participants to interpret and quantify emotions using given annotation formats accurately and collect participants' context without adding much to this load. We recommend that researchers explore more descriptive methods of annotating emotional data, such as using text, audio, or qualitative questioning through chatbots or large language models.
Moreover, creating stimulus-specific questionnaires to comprehend the factors contributing to emotional responses can also be explored. These questionnaires can be pre-designed using insights from pilot trials and qualitative participant interviews. For example, stimuli featuring elements like dogs should include questions about participants' attitudes towards dogs to incorporate potential influences on annotations.
Researchers should also explore experimental design that includes prior participant profiling using either qualitative or survey-based methods. This step enables a comprehensive understanding of participants' personalities and emotional responses in their everyday lives. Capturing participants' personas and attitudes can be achieved by using questionnaires such as standard general psychological well-being tests \cite{qin2018general}, personality traits questionnaires \cite{rammstedt2007measuring}, and experiment-specific reflective questionnaires \cite{10.1145/3334480.3383019}. This approach ensures a more nuanced exploration of individual characteristics, contributing to a richer and more meaningful interpretation of collected data.

\subsubsection{Designing for Participant in the AI Pipeline}

Our findings underscored the critical role of experimental design choices, advocating for collaborative and participant-centred approaches.  Researchers from CSCW and HCI communities should aim to design environments and stimuli that offer realistic psychological experiences \cite{10.1145/3491102.3502129}, thereby minimizing the influence of lab conditions. Moreover, our study highlights the importance of authentic emotional experiences when engaging with stimuli. Currently, much of the research relies on open-source stimuli derived from publicly available videos, music, or films. However, these stimuli may not be specifically curated for training AI models. Future studies could explore the development of stimuli that integrate these elements more effectively for emotional data collection purposes. Another interesting direction would be to explore designing emotional stimuli with humans in the loop. To address the influence of stimulus order, researchers should consider designing emotion-centric experiments tailored to the specific application. This approach contrasts with a one-size-fits-all design, where participants are exposed to a broad spectrum of emotions. Instead, focusing on the application's specific emotional context ensures a more nuanced understanding of participants' responses. Researchers can find inspiration in studies that specifically addressed single emotions such as harmful stress \cite{10.1145/3460418.3479335}, anxiety \cite{10.1145/3544793.3563428}, happiness \cite{10.1145/3329189.3329216} and depression \cite{10.1145/3204949.3208125}. Researchers can look into using qualitative methods and pilot studies with participants to decide the length of stimulus in the case of laboratory settings. Our findings suggested different emotions require different exposure lengths, suggesting a need for deciding the stimulus length based on the emotion category rather than using the same length for all emotions. To handle the complexities arising from mixed emotions, multiple questionnaires in the annotation procedure of emotion data, and reliance on participants' responses, we recommend practitioners employ techniques such as combining discrete and continuous labels \cite{sharma_dataset_2019}, utilizing unsupervised or semi-supervised methods to identify labels from the data, or chat-bot based annotations for more context on emotions \cite{10.1145/3461615.3485670}.
To accommodate participants' biases towards the technology, we suggest future works to design more elaborate familiarization training or acclimatization procedures along with instructions on what is expected from the current level of VR technology.

\subsection{Data-Work supporting Model-Work}
Our research highlights the difficulties associated with collecting physiological emotion data. In this section, we will delve further into another side of the spectrum - designing experiments with less burden on human participants. Currently, AI-based interventions lean heavily on supervised techniques that rely extensively on the availability of accurate data annotations. However, our analysis reveals the inherent challenge of obtaining annotations that precisely correspond with physiological changes. Factors such as the subjectivity of participants, reliance on participant’s interpretations of emotions, and a lack of contextual information in objective labels contribute to the complexity of data collection. These challenges emphasize the need for nuanced approaches and to pursue more accurate and reliable physiological emotion data collection. In the following sections, we will explore opportunities for CSCW, HCI, and AI researchers to design data collection experiments that address these challenges in emotion AI data work.

\subsubsection{Taking Holistic or Application-Centric Approaches}
Collecting physiological emotion data is commonly regarded as a means to gain insights into human emotions. This process typically involves participants either annotating the emotions they experience during the data collection period or being deliberately exposed to emotions of interest as per the study design. Both approaches hinge on the participants accurately identifying and annotating their emotions. Although models trained on existing datasets have demonstrated some level of performance \cite{schmidt_introducing_2018}, their practical implementation remains challenging. This challenge arises from the complexities of accurately translating physiological responses into a reliable and realistic representation of human emotions. Researchers designing experiments should design methodology that approaches data collection from a holistic view wherein contextual data such as participants' characteristics (age, gender, personality), mood, motivation, physical activity, lifestyle, daily routine, location, health conditions, food, and caffeine intake should also be captured, along with emotion annotations \cite{10.1145/3411763.3450367}.
Moreover, data collection and model development are frequently approached as distinct phases in research within AI communities, while CSCW and HCI communities have pointed out the importance of data-centric approaches wherein data collection is also treated as a crucial aspect of AI \cite{10.1145/3491102.3501868, 10.1145/3411764.3445518}. Traditionally, the data-related tasks are carried out independently of considerations for the subsequent model work. This separation results in preprocessing, data aggregation, and data preparation occurring in a later stage \cite{10.1145/3411764.3445518}. To illustrate this compartmentalization in emotion data, consider the process of collecting physiological data in a laboratory setting. In this scenario, participants are intentionally exposed to various positive and negative emotions and scenarios. However, when preparing a model, the data is often categorized into a binary classification of stressful and non-stressful, oversimplifying the richness of the collected information for the purpose of classification. We suggest researchers to approach data collection from an application perspective; for instance, if the application detects harmful stress versus eustress (beneficial stress), the annotation questionnaires should be specifically designed to collect stress information from participants rather than utilizing a standard questionnaire like SAM and PANAS which contains questions on emotions that aren't necessary for future model work. 

\subsubsection{Moving towards Annotation-Free or Minimal Annotation Approaches}
Annotations are critical in AI data, serving as a fundamental component. When annotations fail to represent the underlying data accurately, it may lead to seemingly impressive performance metrics but ultimately produce results that are meaningless or unreliable. Collecting annotations for emotion data is particularly challenging due to the numerous factors that can influence the annotations. To address these challenges, we propose that researchers explore semi-supervised approaches \cite{10.1145/3596246} and weakly supervised methods \cite{8906154} to leverage available data more effectively. In the realm of physiological emotion data collected in real-world settings, there is an opportunity to design experiments that align with circadian rhythms \cite{walker2020circadian}. For instance, capturing physiological data at night could be labeled as representing rest or homeostasis (internal stability), using the temporal nature of physiological changes to inform the labeling process. This approach can potentially help researchers working on emotion data to collect quality data while minimizing the dependence on participants for annotations. 

\section{Conclusion}
Applications for automatic emotion recognition are becoming increasingly common, along with the rise of wearable devices equipped with sensors that can monitor physiological changes. However, collecting well-annotated physiological emotion datasets remains challenging. In this study, we examined 37 participants within VR-based lab settings and tried to point out the challenges from participants' perspectives in the data collection procedure. On analysis we identified a noticeable dichotomy between expressed emotions, physiological changes, and corresponding annotations. Factors such as participants' perceptions, interpretations, and experiment design choices have been shown to influence emotional responses and annotations. We have then reflected upon these factors and how we can navigate these factors for better emotion data collection.
Through this work, we have highlighted the opportunities for future work in designing participant-centric experiments, stimulus, and labeling methods. We have also discussed opportunities for annotation-free or minimal-annotation approaches for the future.

\section{Acknowledgements}

We acknowledge the support of the iHub-Anubhuti-IIITD Foundation, established
under the NM-ICPS scheme of the DST at
IIIT-Delhi. We are also thankful to our participants for their willingness to participate in our study and share their perspectives with us.

\bibliographystyle{ACM-Reference-Format}
\bibliography{sample-base}

\received{January 2024}
\received[revised]{July 2024}
\received[accepted]{October 2024}

\end{document}